# Competition Induced Spontaneous Resonant Annihilation of Turing Pattern

C. Cong and Z.J. Ding[*]

*Key Laboratory of Strongly-Coupled Quantum Matter Physics, Chinese Academy of Sciences;*

*Hefei National Laboratory for Physical Sciences at Microscale and Department of Physics, University of Science and Technology of China, Hefei, Anhui 230026, People's Republic of China;*

*corresponding author: zjding@ustc.edu.cn

**Abstract:** Including an extra reactant in the Gray-Scott reaction-diffusion model, the dynamical competition between different chemical species during the Turing pattern formation can lead to species territory invasion phenomenon among different patterns of respective species. We report a peculiar oscillatory change on respective mass of a 2D pattern under competition condition by numerical simulation. The invaded Turing spot pattern can turn from a steady fading mode into a spontaneous resonant oscillation mode, the firefly lighting mode, which occurs synchronously among nearest neighboring spots, until the largest spot is annihilated finally. Such resonant oscillation behavior for annihilation can continue but in a more chaotic way towards final dying of the pattern by invasion of another pattern. The resonant frequency is found to be intrinsic character of the invaded pattern, depending on its control parameters.

Turing reaction-diffusion equation taking the general form of

$$\partial \boldsymbol{C}/\partial t = \boldsymbol{f}(\boldsymbol{C},\mu) + D\nabla^2 \boldsymbol{C} \qquad (1)$$

is a mathematical modeling of spatial pattern mostly observed in biology and chemistry [1-3], where the vector $\boldsymbol{C}$ represents the chemical concentrations of multi-reactants, $D$ is a diagonal matrix of diffusion coefficients, the function $\boldsymbol{f}$ accounts for all local reactions, and $\mu$ stands for the bifurcation parameter set. Since its experimental verification on chemical reaction [4], the Turing pattern has been extensively studied experimentally and theoretically. Typically, stationary Turing pattern generated with binary reactants has two general types of form, i.e. the spot type which forms a hexagonal array pattern and strip type which form a labyrinthine pattern [5], and blending of the patterns can occur in nature [6]. These patterns can be numerically simulated with the 2D binary Gray-Scott reaction-diffusion equation [7].

Additionally, more complex dynamic patterns have also been observed. One type is breathing mode of single spot for the periodic change of spot size, which may cause spot collapse by control parameter [8]. The breathing mode was studied numerically and analytically with an activator-inhibitor model [9], and it was interpreted as transitions between propagating fronts near a parity breaking front bifurcation [10]. A singular perturbation technique was used to indicate the spot collapsing, expanding and oscillating behavior in a reaction-diffusion system [11]. Another type is the twinkling eye mode [12], which is also an oscillatory Turing spot pattern, originated by the nonsynchronous periodical oscillation of global pattern where each spot has $2\pi/3$ phase shift with its nearest neighboring spots [13]. The mechanism for this twinkling eye mode is attributed to the interaction between a subharmonic Turing mode and an oscillatory mode. Furthermore, by introducing interaction of stationary and oscillatory modes in three coupled layers, which is described by reaction-diffusion equations with five reactants, it has demonstrated a broad family of oscillating Turing-like complex patterns [14].

On the other hand, multi-reactants reaction-diffusion system can be used to explore the species invasion, e.g. the cancer invasion for the spatiotemporal development of tumor tissue [15,16]. Considering the chemical reactions in a water-in-oil microemulsion droplets system [17], the numerical study of extended FitzHugh-Nagumo model for three fundamental reactants has shown back and forth invasion phenomenon due to interaction of two Turing patterns generated in separate compartments [18]. However, the Gray-Scott model is still the prototypical example of spatiotemporal Turing pattern formation in reaction-diffusion systems. In this Letter, we report a novel and interesting behavior in pattern growth process, i.e. the spontaneous resonant annihilation in pattern decay under invasion of another pattern (most remarkably, the spot pattern invaded by the strip pattern), in a simple ternary Gray-Scott model. This study provides a new insight of chemical and biological self-organization behavior under competition environment.

With an additive reactant, the extended Gray-Scott model equation is written as,

$$\begin{cases} \partial u/\partial t = D_u \nabla^2 u - uv^2 - uw^2 + F(1-u); \\ \partial v/\partial t = D_v \nabla^2 v + uv^2 - (F + K_v)v; \\ \partial w/\partial t = D_w \nabla^2 w + uw^2 - (F + K_w)w, \end{cases} \quad (2)$$

where $u$ is density distribution of activator, and $v$ and $w$ are density distributions of two inhibitors. In this expression, there is no direct interaction term between $v$ and $w$, but they interact indirectly with each other through intake of activator. For $v$ and $w$ are similar chemical species, they ought to form patterns belong to the same kind of binary reaction-diffusion model, depending on diffusion coefficients and conversion rates. The third order Runge-Kutta arithmetic is used for 2D numerical solution, while the simulation results have also been confirmed by lattice Boltzmann method. To observe the competition behavior, the two inhibitors are initially set in separate compartments with an isolation between them to prevent from contaction. The compartment defines boundary condition; two types of boundary conditions are considered here, and in each case the outer periphery is a square on which the periodic condition will be applied during competition stage. The isolation is set as a diagonal (equal triangle compartments), centerline (equal rectangle compartments) and a circle (unequal compartments). The square size is $400 \times 400$. The activator is initially put homogeneous ($u = 1$) inside the square except in the inhibitor seed areas ($u = 0.5$).

An inhibitor seed ($v = 0.25$ or $w = 0.25$) is poured into a small area in size of $20 \times 20$ inside each compartment to develop the typical spot- and stripe-patterns by using binary reaction-diffusion equation and by choosing appropriate modeling parameters in the region of Turing instability. When the respective patterns grow up to fill their territories, we remove the isolation and initiate the pattern competition stage with the ternary equation. Due to competition, one pattern will growth further and another one will be declined. Since $v$ and $w$ are symmetrical in Eq. (2), we then name $v$ as victim and $w$ as invader, whose nature is actually determined according to the $D$ and $K$ values.

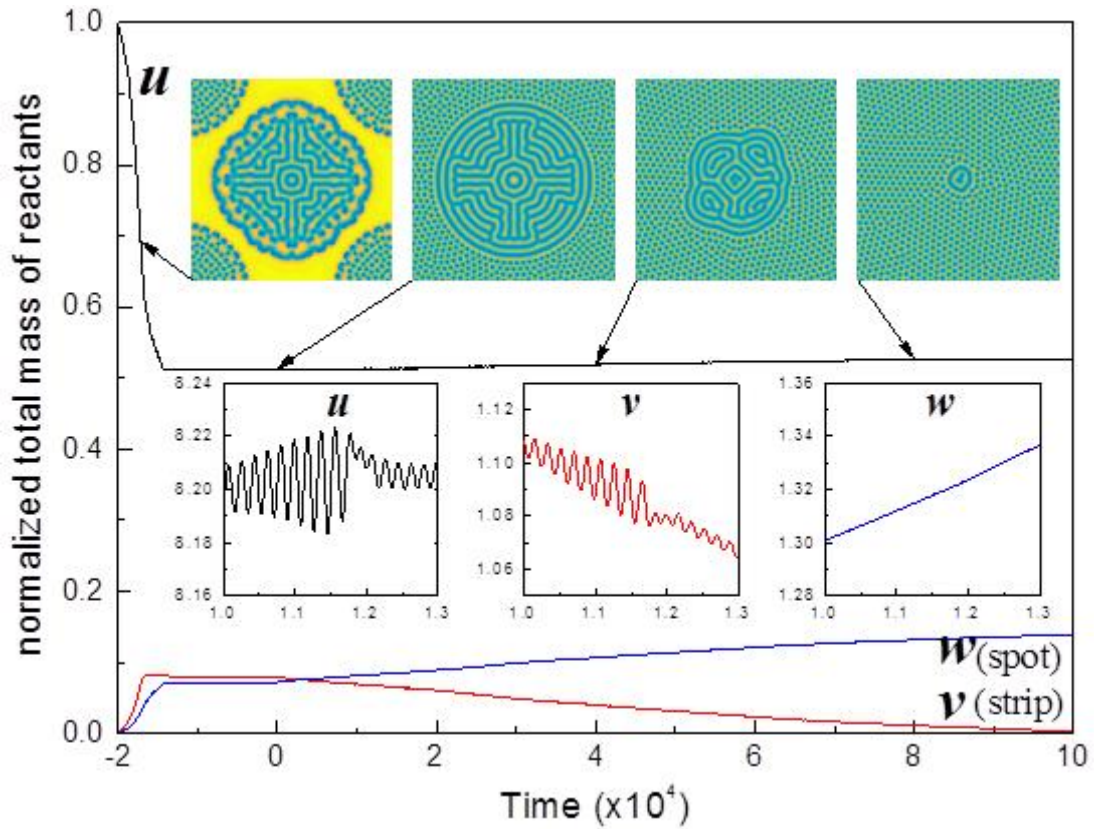

**Fig. 1** The variation of total mass with time for each reactant in the case of invasion of spot pattern into strip pattern. The competition starts from $t = 0$. The bottom insets show the expanded mass curves in a small time-interval, and upper insets are the snapshots of instantaneous $u$-pattern, where the strip (labyrinth) and spot patterns are due to $v$ and $w$ reactants, respectively.

To represent the overall loss/gain of the population of each species, we define the time-varying total mass, $M_z(t) = \iint z(x, y; t) dx dy$, for each reactant $z \in (u, v, w)$, while it is hard to define the territory of a pattern via area. Fig. 1 shows the mass curves for the case of spot pattern invasion into strip pattern, where the parameter set is taken as

$[D_v, K_v, D_w, K_w] = [0.15, 0.055, 0.10, 0.061]$ ($D_u = 0.30$ and $F = 0.026$ are fixed throughout). Even though $v$ grows faster than $w$ in the growing stage ($t < 0$), however, the $v$-pattern (strip) is seen to become weaker under competition and is then invaded by the $w$-pattern (spot), in the competition stage ($t > 0$) to show continuous fading tendency. The invasion behavior is independent of initial condition and boundary condition, but only relates to the parameter set. The mass changes in this case look like steady from the curve, but the magnified intensities show that the invaded $v$-pattern, together with $u$-pattern, has tiny oscillations in a fixed frequency. Because of the massive linked network of the labyrinth pattern and large difference on competitiveness between two patterns, the oscillation amplitude is negligible in this case.

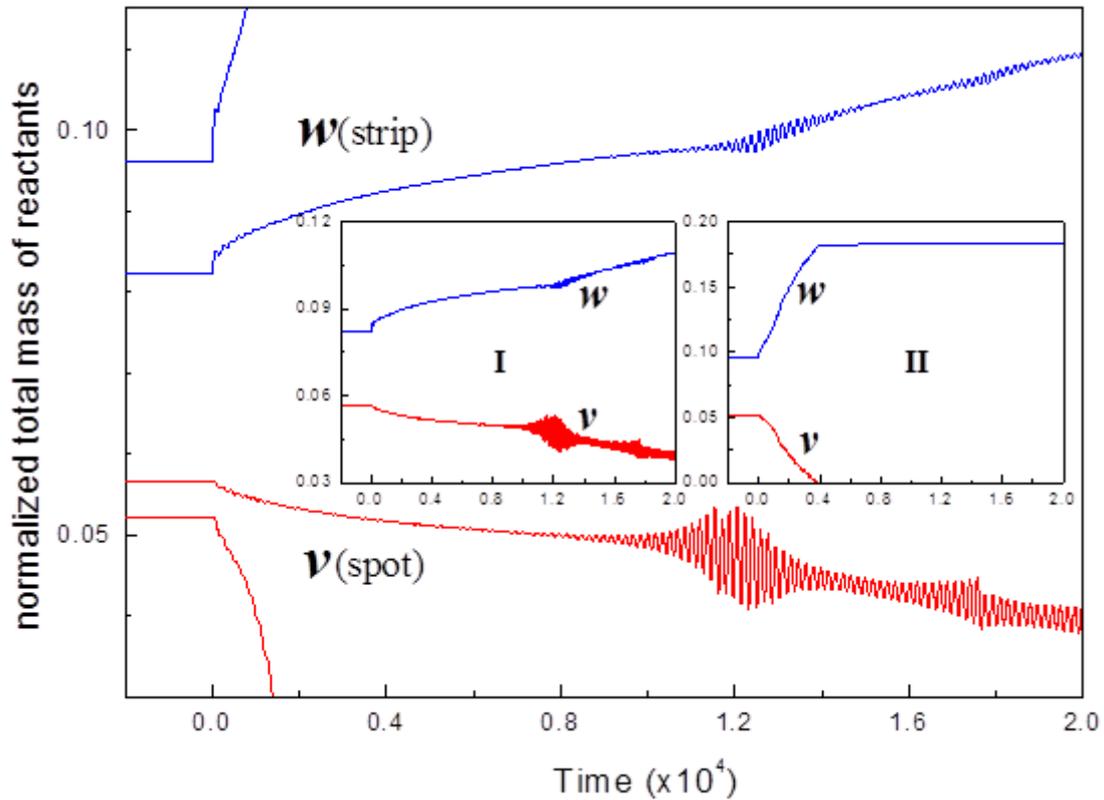

**Fig. 2** The variation of total mass with time for reactants, $v$ and $w$, in the case of invasion of strip pattern into spot pattern for two parameter sets, I and II. The initial triangle boundary condition is used. There is a strong resonant decay of spot pattern in case-I. The insets show the demagnified mass curves in the two cases.

However, in the opposite invasion case of strip pattern into spot pattern, the invaded spot pattern can turn from the trivial steady fading mode into a resonant oscillation annihilation mode, as shown by the total mass curves in Fig. 2 for two parameter sets of [0.13, 0.061, 0.15, 0.055] (case-I) and [0.13, 0.061, 0.10, 0.055] (case-II). In case-II, there is a large difference on the competitiveness between $v$ and $w$, then $w$-pattern (strip) expands the territory and $v$-pattern (spot) vanishes quickly to end the competition. But in case-I, when $w$-pattern invades in a very slow rate, the invasion induces a large resonant oscillation of $v$-pattern (spot) followed by continuous oscillations during its attenuation process. From the pattern and its differentiation map [V1 in supporting materials] in Fig. 3 it can be seen that the oscillation happens in a local area of spot pattern in such a way that several spots are gradually synchronized to change the intensity periodically with enlarging amplitude; such rebound phenomenon finally cause one or several spots to collapse into annihilation. After that nearby spots split by self-replication and move to occupy the left empty space and begin another round of resonant oscillation in a different location. Differentiation map videos in two different initial boundary conditions also produced [V2,V3 in supporting materials]. This collective spot oscillation mode, the firefly lighting mode, on the intensity change of spots is quite similar to the synchronous and resonant lighting of male firefly group in nature. Such resonant annihilation process continues until the $v$-pattern becomes individual spots and single rows separated by the topologically connected $w$-pattern territory so that the nearby spot neighbors are not sufficient to produce spontaneous resonance. In differentiation map, with emphasis on time changing or oscillation behavior, some interesting pattern changes can also be found [F1 in supporting materials]. Globally the changes at different locations are out of phase. Since the initial triangular boundary is symmetric about the diagonal the pattern develops symmetrically in the initial stage of competition; the resonance occurs in the center area about the symmetry axis where a group of spots cooperatively enter into a resonant mode and some of them collapse together. However, due to a slight fluctuation in the system the pattern is gradually distorted into asymmetric spot annihilation. In the circular boundary case, i.e. the initial boundary is made of a circular spot pattern contained inside a square strip pattern, the pattern gradually develops into symmetric about one diagonal; the resonance then is initiated with almost full spots and some spots collapse about the symmetry axis. In later time, the pattern also becomes asymmetric and irregular.

The long-time tails in the decay curve of $v$-pattern are also quite different in the two cases. The spots in case-I dies in a quantum step but not in continuous reduction, i.e. two spots join into one spot; however, they attenuate continuously in an oscillative way in case-II before the complete vanishing of spot pattern. In the limit of $t \to \infty$, the last single spots in the invaded pattern can sustain and becomes stationary, which implies that the ternary equation would be reduced to the binary equation.

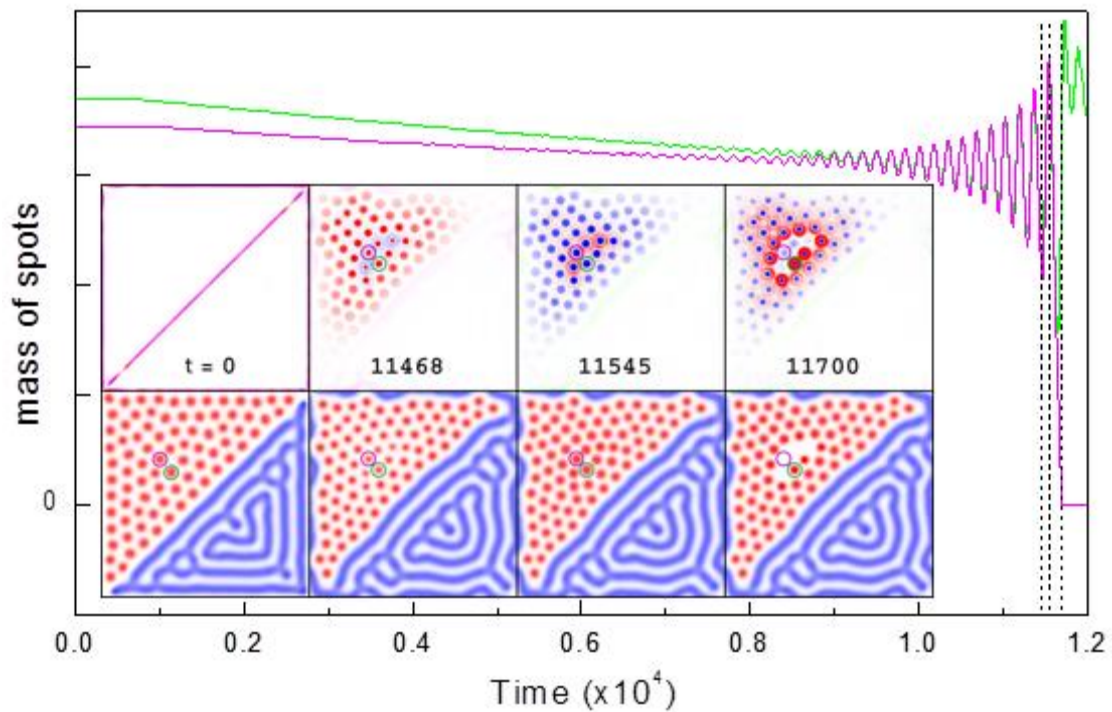

**Fig. 3** The mass variation of two single $v$-spots with time for the case-I in Fig. 2, where the pink curve represents a spot that collapses into annihilation by resonance and the green curve a neighboring spot that continues to oscillate thereafter. The bottom and upper panel insets show respectively the snapshots of mass map and its differentiation map of competition pattern between $v$ (red, spot) and $w$ (blue, strip) at different time instants.

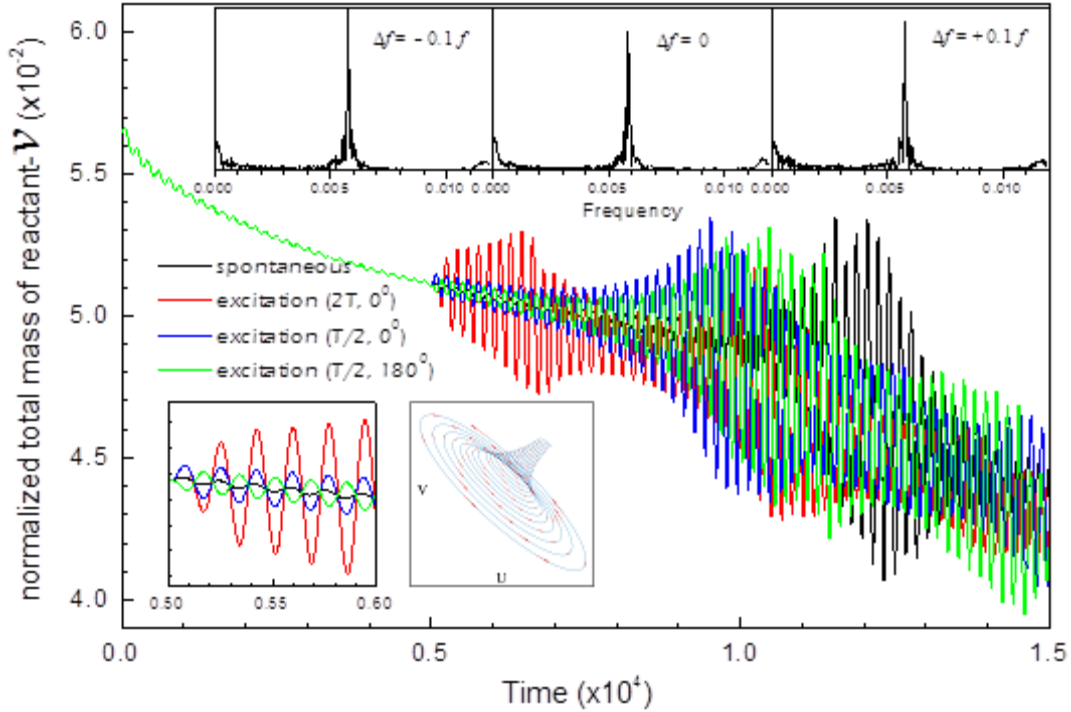

**Fig. 4** The variation of total mass of $v$-reactant with time. The spontaneous resonance (black) is the case-I in Fig. 2, and an excitation of resonance by a sinusoidal perturbance is trigged at $t = 5000$ for time duration $\Delta t$, amplitude $\gamma$, frequency difference $\Delta f$ and phase difference $\varphi$ in Eq. (3): $\Delta t = 2T$, $\Delta f = 0$ and $\varphi = 0$ (red); $\Delta t = T/2$, $\Delta f = 0$ and $\varphi = 0$ (blue); $\Delta t = T/2$, $\Delta f = 0$ and $\varphi = \pi$ (green). The lower left inset shows the onset region of excitation, and the lower right inset is the phase diagram of $uv$ reactants for the spontaneous resonance region. The upper inset panel shows the frequency spectra for excitations with $\Delta t = 2T$ and $\varphi = 0$ but different $\Delta f$.

The frequency spectrum of the mass curve for the $u$-pattern, which oscillates with $v$-pattern in the same frequency (as seen in the phase diagram in Fig. 4) but with larger amplitude, shows that there is only a single main peak of oscillation ($f = 0.0584$) while the small subsidiaries are due to the complex transit process after spot annihilation (Fig. 4). Although the frequency is single valued which is determined only by control parameter set, the onset and oscillation processes is found to strongly depend on initial boundary condition. This fact implies that, this spontaneous oscillation is most likely resulted from a transient local fluctuation in activator intensity under the slow but continuous invasion of an invader pattern. The stochastic resonance phenomenon has been well known in a wide range of dynamical systems, in which a weak periodic stimulus entrains a large-scale environmental oscillation [19]. To understand the mechanism of resonance and the role of fluctuation, an external noise can be introduced into the system [20]. Here, we introduce an external sinusoidal perturbance [18] in the first equation in Eq. (2):

$$\partial u/\partial t = D_u \nabla^2 u - uv^2(1 + \gamma \sin(\omega' t + \varphi)) - uw^2 + F(1 - u) \qquad (3)$$

where $\omega' = 2\pi(f + \Delta f)$ is excitation angular frequency, $\Delta f$ is frequency change with respect to the resonant frequency $f$, and $\varphi$ is phase shift. The excitation source is applied to a circle area of radius of 50 away from the symmetry axis with a very small amplitude of $\gamma = 0.01$. At the location of the applied perturbance the resonance is quickly excited as expected to demonstrate a shorter resonance lifetime as compared with the spontaneous resonance; the followed resonant decay process also changes drastically depending on the duration of excitation period $\Delta t$ and phase shift (Fig. 4). For $\Delta t = 2T = 2/f$ and $\varphi = 0$, a quick resonance is generated just within the excitation duration [V10 in supporting materials]. For $\Delta t = T/2$ and $\varphi = \pi$, i.e. in the case of the excitation lasted only for the half period of spontaneous oscillation and in opposite oscillation direction, the resonance can also be excited and in reverse oscillation phase according to the excitation phase (lower left inset in Fig. 4). Different excitation frequencies, $\Delta f = -0.1f$ and $\Delta f = 0.1f$, are also applied to see if the resonance frequency can be modulated. The frequency spectra in the insets show that the excited oscillation frequency ($f = 0.0577$) agrees almost with the resonant frequency within 1% of difference no matter what the excitation source is. Furthermore, if a stronger stimulus is applied ($\gamma = 0.05$, $\Delta t = 2T$, $\varphi = 0$ and $\Delta f = 0$), the resonance is more quickly excited, leading to a faster damping of pattern. In contrast to the resonant annihilation for ternary reactants, when such a stimulus ($\gamma = 0.01$, $\Delta t = 2T$, $\varphi = 0$ and $\Delta f = 0$) is applied to binary reactants and a pure stationary $v$-pattern produced by a binary equation, the oscillation attenuates without resonance annihilation and the pattern becomes stationary again. Therefore, the resonance is the direct result induced by the invasion of invader pattern. Analytical and numerical studies on 1D binary Gray-Scott model have shown the two-spike solution has competition instability and oscillatory instability, and the later synchronizes the amplitudes of the spikes and leads to annihilation of spike [21,22]. Such a synchronized oscillation may explain the origin of the intrinsic resonance without invasion, and hence the invader plays a role to induce instability.

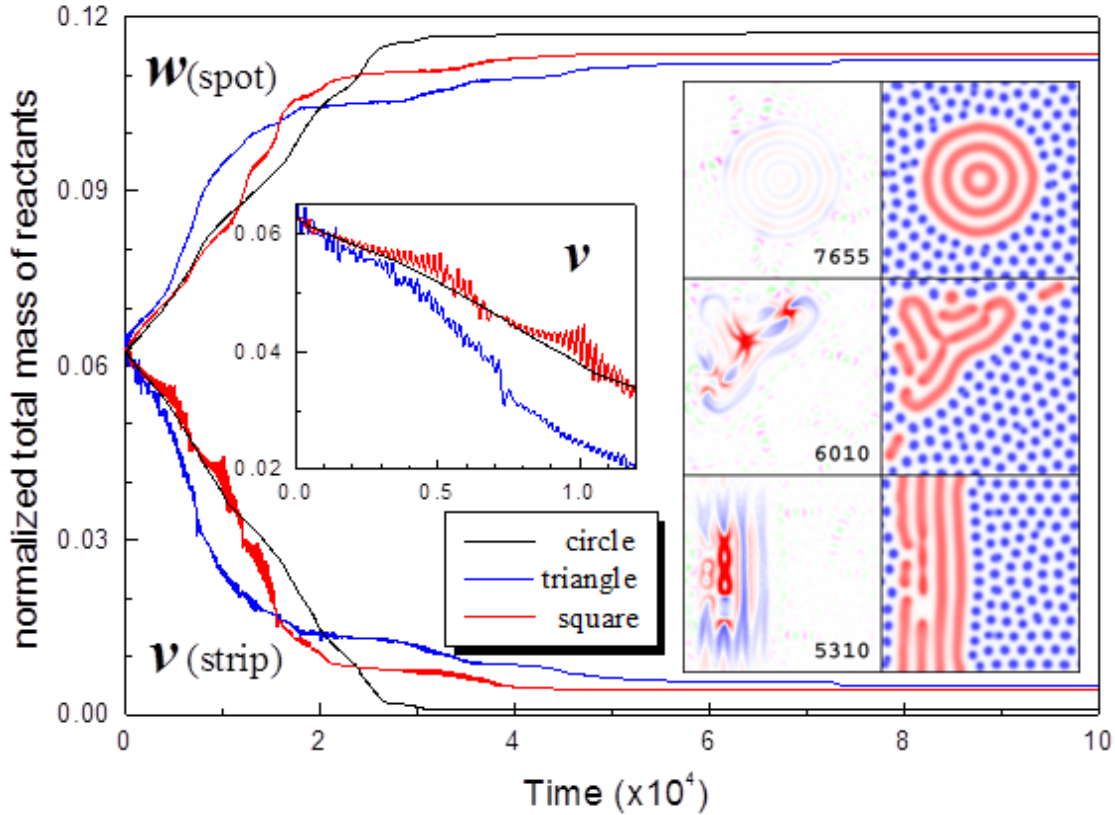

**Fig. 5** The variation of total mass with time for reactants, $v$ and $w$, in the case of invasion of spot pattern into strip pattern for three initial boundary conditions. The left inset shows the magnified $v$-curves. The right and left panels are instaneous pattern and its differentiation map for different initial boundary conditions: circle (top), triangle (middle) and square (bottom).

In fact, in the case of spot pattern invasion into strip pattern the oscillation can also happen but in a rather complex way, when the two patterns have close competitiveness, e.g. for parameter set of [0.20, 0.055, 0.13, 0.061]. The oscillation location in the strip pattern may happen either at terminal joints of strips or a short strip, which can be visualized through the pattern and its differentiation map [V4,V5,V6 in supporting materials]. The resonance is less easily to be generated spontaneously among strips due to extensive labyrinth network in comparison with the spot resonance case. Fig. 5 indicates that the oscillation amplitude is rather irregular in time, and, it is very sensitive to the initial boundary condition. In the circular boundary case, because the strips are all in circular shape and totally uniformly connected, there is no oscillation found; the outer circular strips continuously shrink and the inner most circular strip becomes a spot and then absorbed by the outer strip later. In the triangle boundary case, there is only slight irregular oscillation but without resonance. The largest oscillation as quasi-resonance is found for the rectangular boundary condition; the quasi-resonance among parallel strips breaks up a long strip into several shorter ones. In all the cases, the final fate of strips becomes stationary single spots.

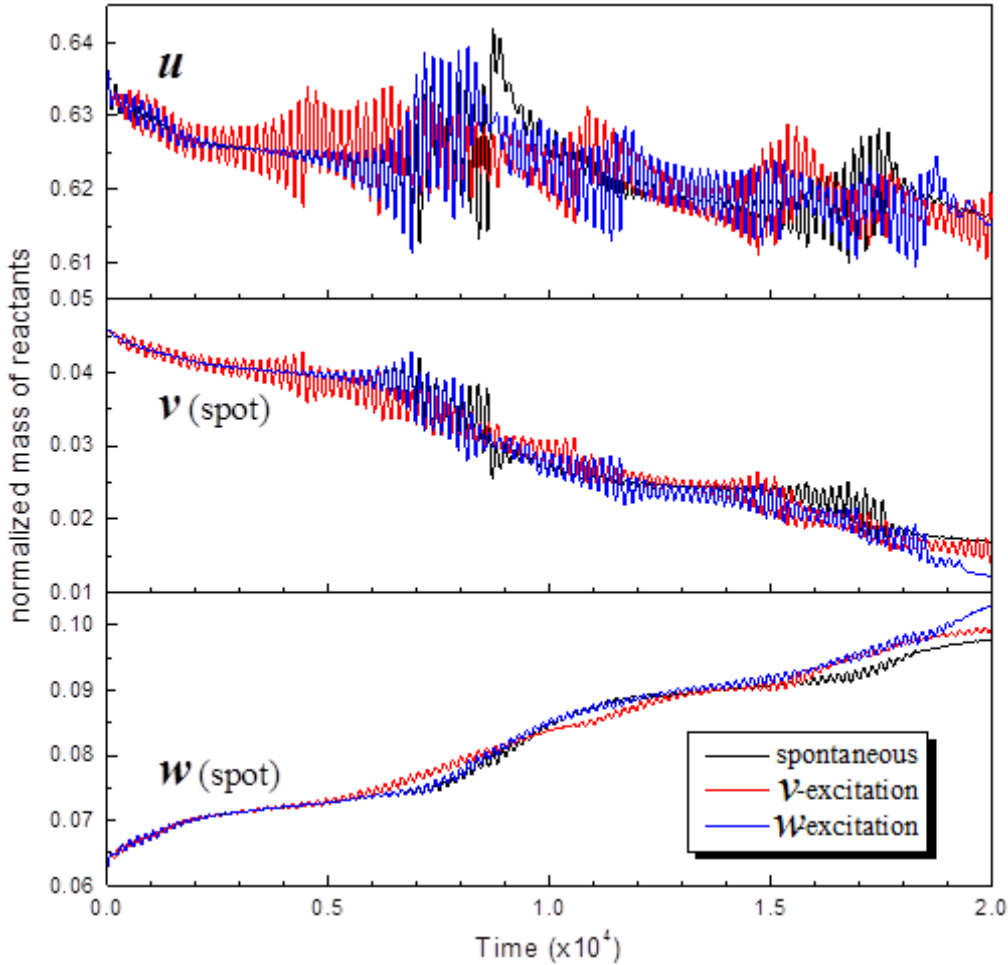

**Fig. 6** The variation of total mass with time for reactants, $u$, $v$ and $w$, in the case of invasion of spot $w$-pattern into spot $v$-pattern under the circular initial boundary condition. The spontaneous resonance is compared with the excitations in the $v$- and $w$-patterns.

Another remarkable resonance is found to occur for the spot pattern invasion into spot pattern [V7,V8,V9 in supporting materials], Fig. 6 illustrate the mass curves for the parameter set of [0.15, 0.061, 0.13, 0.061]. Not only the $v$-pattern but the $u$-pattern demonstrates more prominent oscillation and successive resonances in time. The invader spot pattern expands the territory by spot self-replication near the boundary of the two patterns to increase population while keeping the wavelength of the pattern. Global spots other than local spots can be cooperatively involved in the spontaneous resonance, and hence lead to great oscillation amplitude. Once the invaded spots of the $v$-pattern are annihilated after resonance the surrounding spots of $w$-pattern splits and then invade the space. Periodic perturbance ($\gamma = 0.01$, $\Delta t = 2T$, $\varphi = 0$ and $\Delta f = 0$) is then introduced into both the $v$- and $w$- spot patterns, at $t = 0$, to see the difference on excitation result [V11,V12 in supporting materials]. The excitation effect in the $v$-pattern is found similar to the case of strip pattern invasion into spot pattern. However, the excitation in the $w$-pattern can also result in a long-term resonance effect to be different from the pure spontaneous resonance. This fact indicates the resonant annihilation of spot is sensitive to many environmental perturbance sources; but, the giant resonance can only occur in the invaded pattern while the invader pattern only presents a damping behavior.

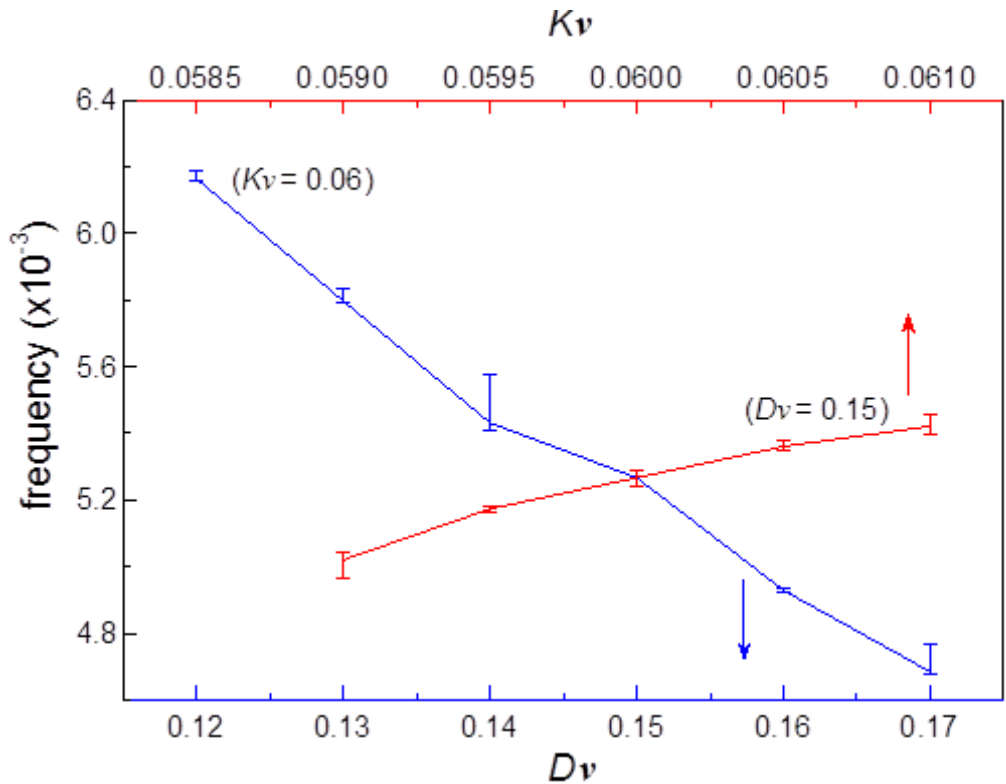

**Fig. 7** Dependence of resonant frequency on control parameters, $D_v$ (when $K_v$ is fixed) and $K_v$ (when $D_v$ is fixed), of the $v$-pattern obtained for the invasion of strip pattern into spot pattern. The error is due to the peak width in a Fourier spectrum.

The simulation results have shown that the resonant frequency is independent of control parameters of invader $w$-pattern and it is only the intrinsic property of the invaded $v$-pattern. Fig. 7 indicates that a relationship between frequency and control parameters exists, $f \sim K_v/D_v$, obtained for the case of strip pattern invasion into spot pattern.

Therefore, in conclusion, the observed firefly mode is a cooperative resonance mode of the invaded pattern of the nonlinear Gray-Scott system driven by the fluctuation under the continuous change of boundary of the invader pattern. This boundary change plays a role of external force to induce the fluctuation and instability.

This work was supported by the National Natural Science Foundation of China (No. 11574289) and Special Program for Applied Research on Super Computation of the NSFC-Guangdong Joint Fund (2nd phase). We thank supercomputing center of USTC for the support of parallel computing.

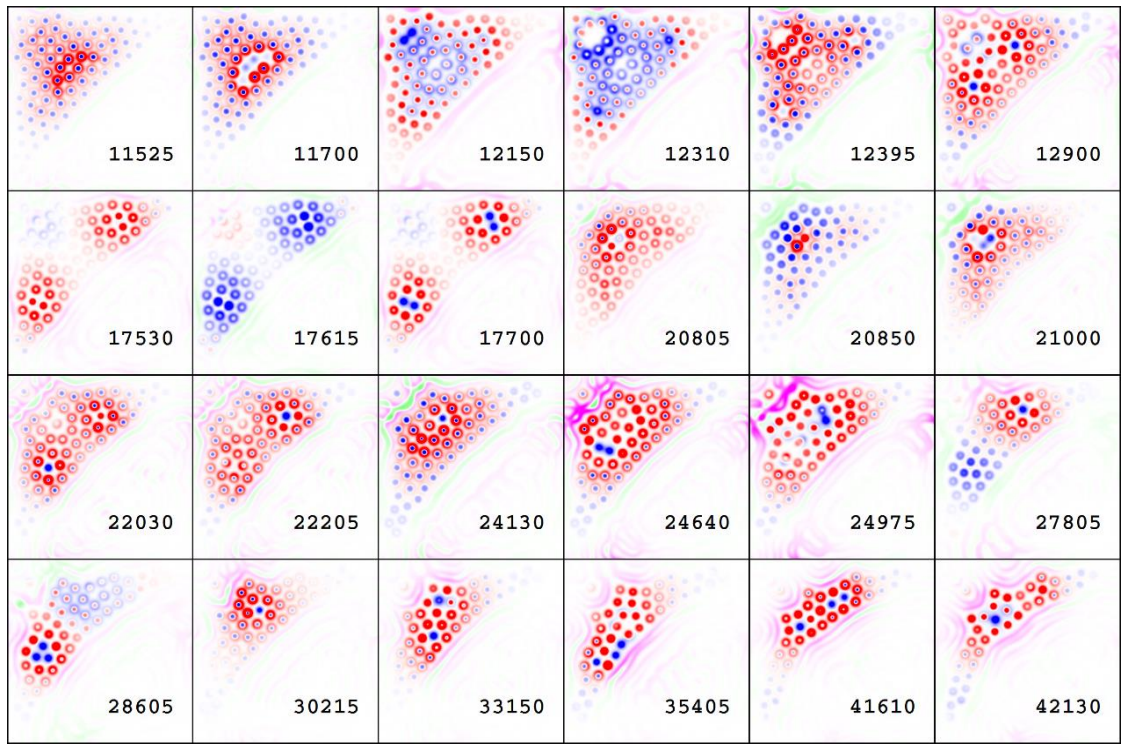

F1. The snapshots of the differentiation map for the competition process at different time instants in the case-I in Fig. 2.